\documentclass[%
reprint,
superscriptaddress,
showpacs,preprintnumbers,
nofootinbib,
 amsmath,amssymb,
floatfix,
]{revtex4-1}
\usepackage[colorlinks=true,
urlcolor=darkraspberry,
linkcolor=darkraspberry,
citecolor=darkraspberry,
linktocpage=true,
pdfproducer=medialab,
pdfa=true,
anchorcolor=blue]{hyperref}
\usepackage{xcolor}
\definecolor{cambridgeblue}{rgb}{0.64, 0.76, 0.68}
\definecolor{darkraspberry}{rgb}{0.53, 0.15, 0.34}
\usepackage{graphicx}
\usepackage{dcolumn}
\usepackage{bm}
\usepackage{amssymb}
\usepackage{placeins}
\usepackage{comment}
\usepackage{subcaption}

\usepackage{tikz}
\usepackage{etoolbox}
\usepackage{orcidlink}
\usepackage{float}
\usepackage[utf8]{inputenc} 
\usepackage{xcolor}
\usepackage{hyperref}
\usepackage{fancyhdr}
\usepackage[normalem]{ulem} 

\begin{document}

\begin{titlepage}

	\title{``Niñas Atómicas" (Atomic Girls): An initiative that generates opportunities for young girls in STEM}
	\author{Giovanna Cottin\,\orcidlink{0000-0002-5308-5808}}
	\email{gfcottin@uc.cl}
	\author{Francisca Garay\,\orcidlink{0000-0002-6670-1104}}

	\affiliation{Instituto de F\'isica, Pontificia Universidad Cat\'olica de Chile, Avenida Vicu\~{n}a Mackenna 4860, Santiago, Chile}
\affiliation{Millennium Institute for Subatomic Physics at the High Energy Frontier (SAPHIR), \\ Fern\'andez Concha 700, Santiago, Chile}

	\date{\today}
	\begin{center}
   \includegraphics[width=0.3\textwidth]{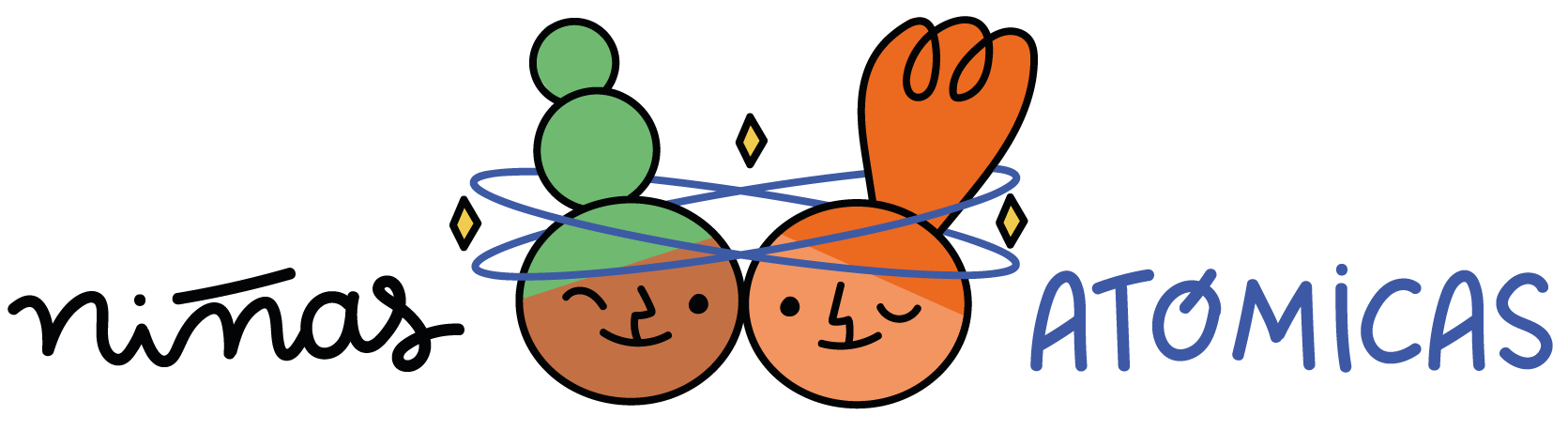}  
\end{center}

\begin{abstract}
We report on an initiative that seeks to encourage high school girls to develop critical thinking and transferable skills widely used in scientific work, as well as to generate a concrete space of opportunities for girls to experience how real science is done. 
Our ``Niñas Atómicas" workshop combines the teaching of particle physics, electronics, programming and scientific methodology through building and operating a dedicated experiment:  a muon counter. Girls from all over Chile can apply to this workshop, where every year they are guided by female scientists for two weeks. We report on the contents and methodology of our workshop and provide details on how to build the muon detector. We report results on muon flux and proper lifetime, two muon properties which can be extracted from the data collected by the girls with the muon detectors they built themselves. Insights into the girl's experiences during the 2024 and 2025 editions of the workshop are also detailed, with the aim to contribute to the wider physics education research and outreach communities.

\end{abstract}

\maketitle
\end{titlepage}

\section{Introduction}
\label{sec:Intro}

Science is everywhere. However, the education of it and the opportunities to experience it are not universally available and gender inequalities exist. According to UNESCO~\cite{UNESCO}, only $35\%$ of higher-education students that are enrolled in subjects related to science, technology, engineering and mathematics (STEM) are female. Globally, only $28\%$ of science researchers are women.

Particularly in Chile, gender scoring gaps at the high-school level in mathematics~\cite{Radovic2018,PerezMejias2021} may correlate with gender stereotypes~\cite{DelRio2016}, lack of motivation, confidence or limited access to resources and educational opportunities~\cite{TIMMS2019}. In this context, our science workshop ``Niñas Atómicas" (Atomic Girls)~\cite{Atomicas} aims to provide a concrete space of opportunities for high-school girls to experience how real science is done\footnote{They are taught by scientists, they built their own experiment, take data with it, analyze it, come up with a scientific question, and then report on their research results. More details in section~\ref{sec:TheWorkshop}.}. Our objectives are twofold: i) foster critical thinking and transferable scientific skills useful in our modern world, such as programming or a basic understanding of electronic systems, and ii) that this experience aids in their decision-making process by giving them a concrete understanding of what is science and how science is done, before choosing a major\footnote{Having completed STEM related courses at the high-school level helps define a ``STEM-readiness"~\cite{ChoiceSTEM} metric, which can impact the chances of entering a STEM program.}. The core transferable scientific skill we are targeting is the ability to interpret numerical data through mathematical reasoning, within a scientific context, in order to perform a quantitative analysis.

Our workshop revolves around a hands-on experiment that counts muons using a dedicated muon detector. In section~\ref{sec:Exp} we motivate and describe the experiment. In section~\ref{sec:TheWorkshop}, we describe the contents and methodology of the workshop. In section~\ref{sec:results} we present results on muon properties based on the data the girls took after building their detectors, as well as results addressing the perception the girls had after finishing the workshop. We conclude in section~\ref{sec:summary} hoping that the ideas our initiative brings may be replicated or inspire others alike.

\section{The muon experiment}
\label{sec:Exp}

{\bf{Why muons?: }} Muons are elementary particles belonging to the second generation of leptons in the Standard Model of Particle Physics~\cite{Thomson:2013zua}. They are particles about 200 times heavier than electrons, and can penetrate deeply through matter without losing too much energy. Muons are produced in the decays of charged mesons when cosmic rays interact with our atmosphere. They have a proper lifetime close to $2.2\mu$s~\cite{PDGMuonLifetime}, which (thanks to Einstein's special relativity), allows them to travel several kilometers before reaching the Earth's surface. They are the most abundant electrically charged particles at sea level~\cite{PDG2011CosmicRays}. Roughly, one muon per second transverses the surface area of a hand. Therefore, if we build a small detector, and place it at different altitudes on the surface of the Earth, we can measure them with different fluxes. 

\vspace{0.5cm}

\begin{figure}
\includegraphics[width=0.5\textwidth]{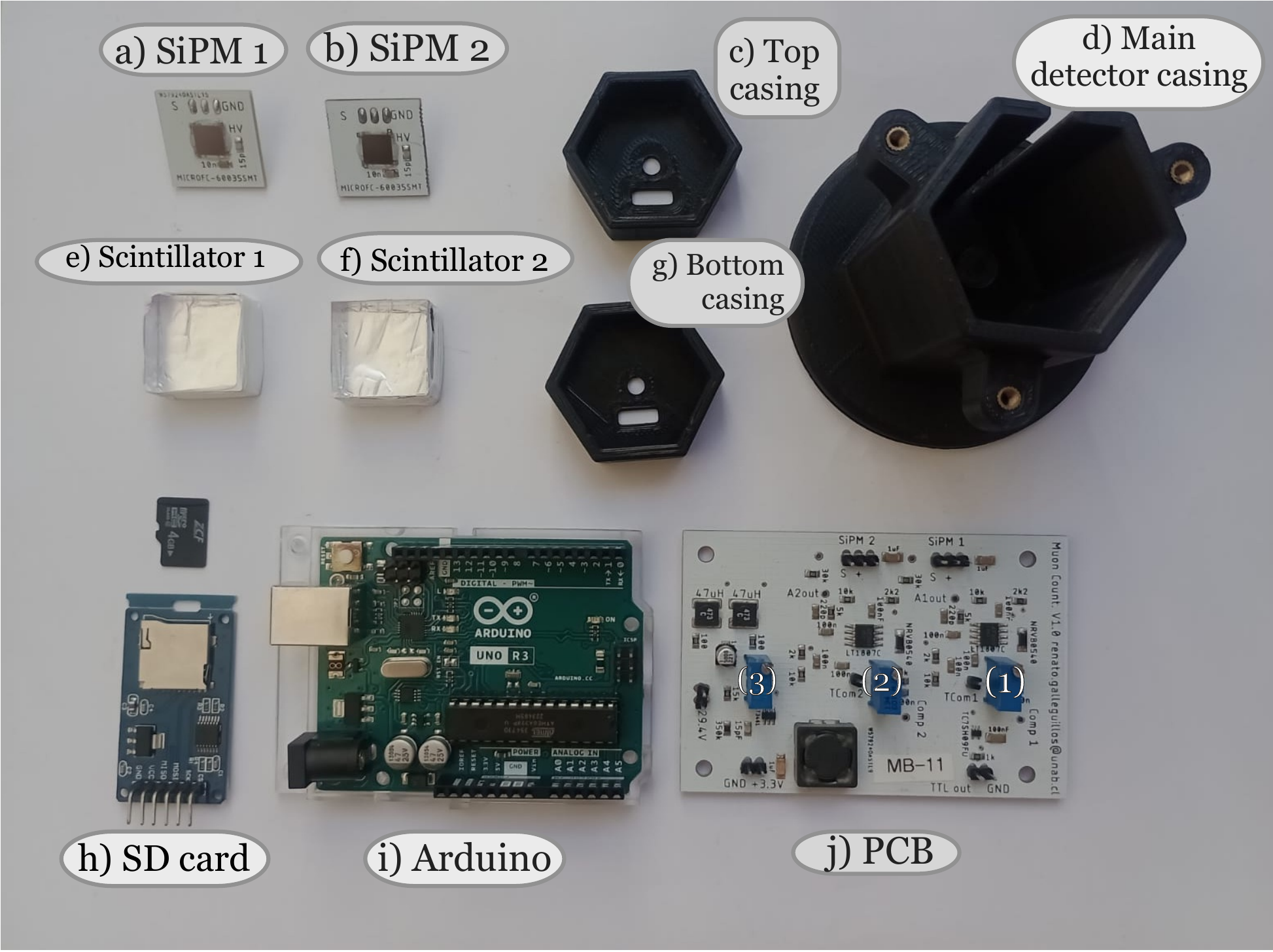}
\caption{Components of our muon detector. The SiPMs are shown in a) and b). The hexagonal casing in c) is placed on top of the main detector casing shown in d). The SiPMs support the scintillators in e) and f), which are wrapped in aluminum foil and are placed on top of them. The bottom detector casing is shown in g). The hexagonal casings enclose the SiPMs when the detector is assembled. The SD card reader in shown in h), which connects to the Arduino in i). These connect to the PCB shown in j).  Figure adapted from a photograph taken by Ignacio García.}
\label{fig:detectorComponents}
\end{figure}

{\bf{Experimental setup:}} The muon detector we use in our workshop includes two plastic scintillators, each equipped with a light-sensing silicon photomultiplier (SiPM). The scintillators are wrapped in a reflective foil to concentrate the light emission to the SiPM. The detector also includes two hexagonal casings and a main case, made from plastic rolls (filament) of PLA plastic printed with a standard 3D printer. It is important that the filament is black, as other colors can sometimes let light seep through. The print quality should also be good to make the piece as dense as possible, or alternatively, the infill parameters can be increased so that the piece acts as a black box\footnote{With printers that provide a high-quality finish, this adjustment is often unnecessary.}. This is so the SiPMs are covered in a dark, non-reflective surface in order to reduce external light contamination to the muon signal. A scintillating material has the property of (re-)emitting energy in the form of photons when a charged particle (as a muon) deposits energy in it. These photons are then processed by the SiPMs, triggering a Geiger discharge and creating a measurable current. The current then goes through a custom-made\footnote{Owned and developed by the Center for Theoretical and Experimental Particle Physics (CTEPP) at Universidad Andrés Bello.} printed circuit board (PCB), which is the physical platform that tailors and processes the signal, and where the electronic components are installed. The signal from the PCB then enters an Arduino UNO from where we can record the data to an SD card. The Arduino is then connected to a computer, and through the Arduino IDE Software~\cite{Arduino}, a count of the detected muons will be recorded, together with a time stamp at which each muon was detected. In our muon detector design, we require a coincidence between two detector channels to declare a valid muon event. Each detector’s analog signal is passed through a discriminator (comparator) to produce a clean logic pulse. These pulses are inputs to a fast AND gate, which outputs a logic “high” only if both inputs go high within a short time window (on the order of nanoseconds). This coincidence requirement suppresses noise and random single-detector triggers: only when the two SiMPs fire nearly simultaneously is a muon detection registered. The data is saved to a {\texttt{.txt}} file for further analysis. The components of the muon experiment are shown in Fig.~\ref{fig:detectorComponents}.

\section{The workshop}
\label{sec:TheWorkshop}

Our ``Niñas Atómicas" workshop has been held every Chilean winter since 2022, lasting two weeks. Girls from all regions of the country can apply online. This article presents the results from 50 participants selected in 2024 (21 selected from 70 applicants) and 2025 (29 selected from 257 applicants). The selection process and workshop methodology implemented was the same for the 2024 and 2025 editions.  The criteria to participate was to complete an online application\footnote{The online application form was advertised in social media, and was embedded within the Qualtrics Software~\cite{Qualtrics}. They were asked to provide their socio-demographics information, to answer questions on their scientific motivations as well as technical questions intended to measure their scientific skills and critical thinking. A separate article is currently in preparation based on the application instrument we created (see Ref.~\cite{EstudioImpacto}).}  as well as being between 15 and 16 years of age. Selection was random among all girls that satisfied the criteria.

The workshop lasts for two weeks in hybrid mode during the school winter holidays in Chile. We start with online classes on particle physics, electronics, and introduce scientific methodology. The girls also travel to the physics laboratories in the San Joaquín campus at Pontificia Universidad Católica de Chile in Santiago for two days, where they assemble their muon detectors and take data in groups of $\sim 5$ girls, each lead by a tutor. Our tutors are all female undergraduate and graduate students in physics, previously trained in the contents of the workshop and the muon detectors. Each tutor has a fixed group of girls and is responsible for guiding them throughout the workshop, evidencing their achievements and difficulties. After taking their data in person and formulating a concrete scientific question about muons, they go back home for the second week of the workshop, where we go back to online lessons on programming and data analysis. We also allocate time for them to report on their research results and their answers to a specific scientific question.

In what follows, we detail the specific sessions of the workshop and what we teach. Throughout all sessions, in selected moments, we ask open questions for the girls to answer in order to diagnose their previous knowledge and to understand their level of comprehension of each session. They can also ask questions anytime.
\vspace{0.5cm}

{\bf{Welcome:}} We start with a $45$ min. long online session where we introduce the workshop objectives, timetable, clarify the high level of commitment it is expected from them\footnote{A $90\%$ of attendance to the sessions as well as handling to their tutors their research report at the end of the workshop were necessary for receiving a diploma.} and present our team: two scientists in particle physics and university professors (the founders of this initiative and authors of this article) that teach the particle physics, scientific methodology and programming lessons, an electronics teacher and 6 tutors (all university students in physics). We also present our administrative and logistics team at the SAPHIR Millenium Institute, which is the main organization supporting the workshop. We leave about 15 minutes for questions from the girls. The groups of girls are assigned to each tutor.

\vspace{0.5cm}

{\bf{Particle physics:}} The level of knowledge in physics we have seen among students is very diverse.  Most of them are familiar with atomic models, but have never heard of elementary particles nor understand why they are necessary to comprehend the universe we live in. In three different online sessions, each lasting about an hour, we teach them fundamental principles underling the nature of elementary particles, focusing on muons. The girls attend these sessions synchronously, listening to the lectures and following the material. At specific points during the sessions, they are invited to answer targeted questions, enriching an interactive discussion on the contents presented. At the end of each session, they are also encouraged to ask questions, further promoting discussion and engagement.

In the first module, three main concepts are explained: matter, scales and interactions. We start by defining what matter is. We foster their curiosity and explain we only know about $5\%$ of the matter content of the universe, as $95\%$ of it corresponds to dark matter and dark energy~\cite{Planck:2013oqw}. We highlight that scientific research is important if we want to understand that big $95\%$, and highlight that the workshop focuses on the known baryonic matter, summarized in the Standard Model of Particle Physics~\cite{Thomson:2013zua}. We introduce the subatomic scales and the physics that governs these scales: special relativity and quantum mechanics. After introducing these contents, the girls are invited to provide their own answers to questions such as:{\textit{``What are we and the things around us made of?"}}, enhancing the depth of the discussion.

We then explain that subatomic particles interact by exchanging other particles (i.e. force carriers as gluons, photons and the $W$ and $Z$ weak bosons), and so we can see them if we can somehow evidence their interactions. We explain the discovery of the electron by J.J. Thompson~\cite{JJThomsonPEdu}, the first elementary particle to be discovered, and show them real pictures from this experiment taken at the Cavendish Laboratory. We finish by highlighting the main properties of the 17 current known particles (and their antiparticles) and known forces that determine how particles interact: the electromagnetic force which puts atoms together, the strong nuclear force that puts nuclei together, and the weak force that tears nuclei apart, wrapping up the concepts of matter, scales and interactions. The students then respond to questions such as:{\textit{``How things in the universe interact?"}}, articulating their own explanations.

In the second session we explain what a muon is, how it is created and what are its main properties. We start by emphasizing that not all particles live inside atoms, and tell them how muons are created in cosmic ray air showers and how they were discovered by Carl Anderson and Seth Neddermeyer~\cite{PhysRev.51.884}. We then explain the properties of the muon: its large mass compared to the electron, its instability via the weak force, its proper lifetime and its large lab frame decay distance, allowing us to detect them. The students through the session provide their own answers to questions such as:{\textit{``How can a muon falling from the sky interact with matter?"}}.

The last particle physics session starts by summarizing the Standard Model of Particle Physics. We try to explain broadly the equation of the Standard Model, and clarify how from it we can identify the different particle interactions (See Ref.~\cite{SMMug} for a pedagogical explanation of the equation). We tell them the role of the Large Hadron Collider at CERN~\cite{CERN} in completing the Standard Model with the discovery of the Higgs boson~\cite{ATLAS:2012yve,CMS:2012qbp}. We further emphasize on the synergy between theory and experiment in the hunt for new particles. We emphasize that the science connected to CERN is being carried out by people from diverse countries and backgrounds with one main objective: to unveil nature's most profound secrets at the subatomic scales. We finish the session with concrete applications particles and muons have beyond particle physics, such as muon mapping~\cite{NatureMuons,NatureMuons2}, in order to exemplify how the science that emerges from particle physics research can have impact in other fields.

\vspace{0.5cm}

{\bf{Electronics:}} Three online sessions lasting about one hour each are devoted to electronics. Aligned with the pedagogical approach of the particle physics sessions, the girls interact with the class by making questions at any time and addressing specific questions through the lessons. During the first session, we define electricity, current, voltage and different types of materials and their susceptibility to conduct electricity (conductors, insulators and semiconductors)\footnote{For an educational capsule aimed at high-school students (in Spanish) on what is electricity and how to generate it, see Ref.~\cite{cottin_curriculumnacional}.}. We also explain different types of electric components. These include power sources, switches, capacitors, resistors, inductors and scintillating materials. Discussions around questions such as {\textit{``What is electricity, and where can you find it in your home?"}}, promote the exchange.

In the second session we continue discussing electrical components, now focusing on the components the girls will use to assemble their muon detectors: diodes, photodiodes, and silicon photomultipliers (SiPM). We finish the second session by detailing the different types of circuits (in series and in parallel) and two instruments to perform active (as voltage) or passive (as resistance) measurements: a multimeter and an oscilloscope. A specific circuit example is presented at the end of this session, where the girls respond to questions such as: {\textit{``If I want to measure the voltage across the resistor in the circuit, how should the measuring device be connected?''}}

\begin{figure}
\includegraphics[width=0.45\textwidth]{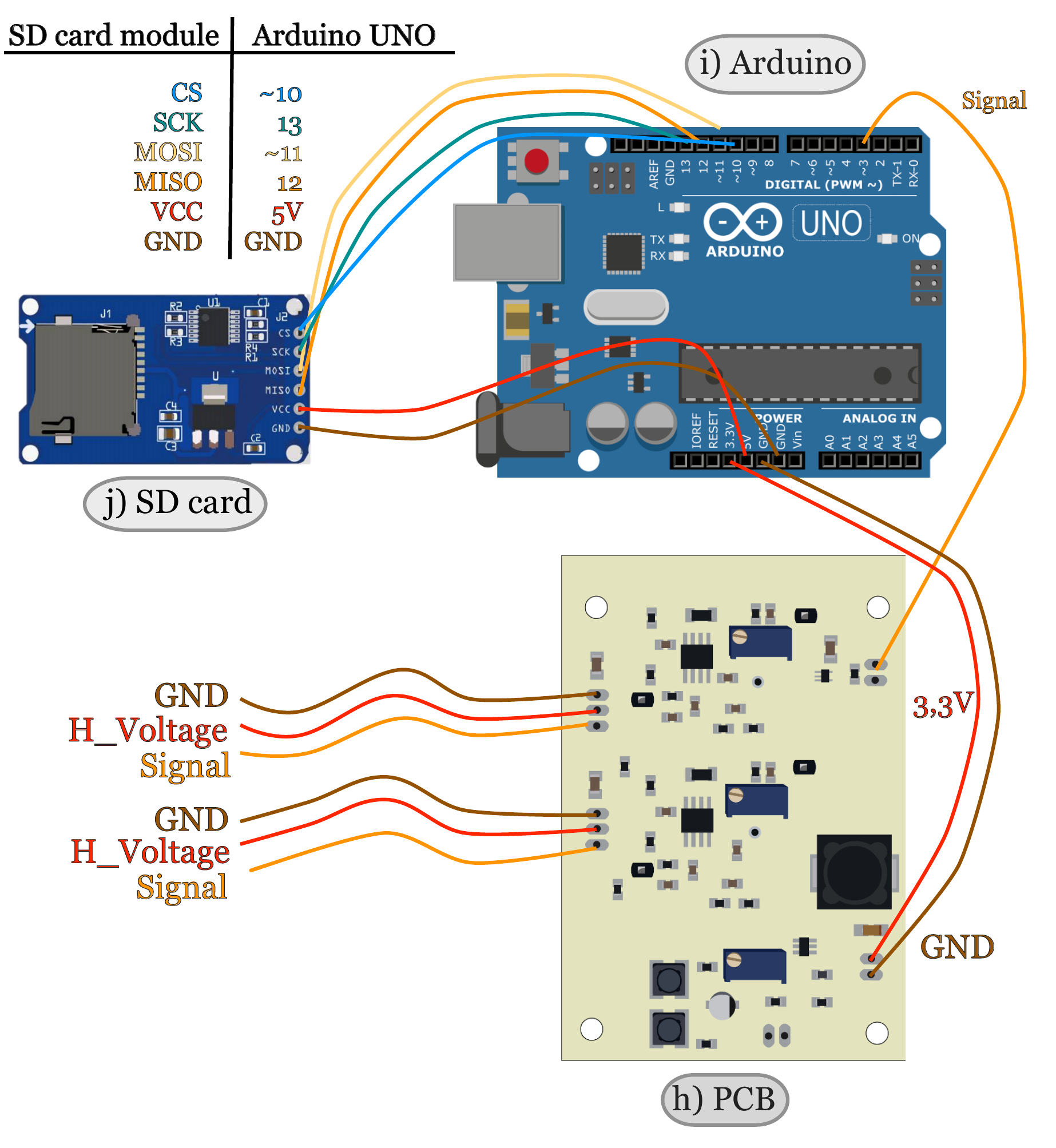}
\caption{Diagram showing the electrical connections of our muon detector. Detailed instructions can be found in~\cite{AtomicasGithub}. We thank Roberto Pinto for this diagram.}
\label{fig:Electronics}
\end{figure}

The third and final electronics session is devoted to explain the characterization of the path a muon will undertake as it travels through the muon detector. The different detector components in Figure~\ref{fig:detectorComponents} are explained highlighting their relevance in making the experiment work.  The different connections of the SiMPs to the PCB are detailed, as well as how the signal is processed. When there is a coincidence of two nearly simultaneous pulses, the PCB counts a signal. This means a muon passed through both scintillators, which is when a muon is detected. Furthermore, the Arduino is explained detailing how we can save the data with its SD card reader module. The electrical connections detailed in Figure~\ref{fig:Electronics} are explained, so they know beforehand what they will need to connect before assembling the detectors themselves. We finish by explaining the role of potentiometers to avoid failures in the PCB and to make sure we feed the detector with the adequate current and voltage.

\vspace{0.5cm}

{\bf{Hands-on detector assembly:}} The girls (together with one parent or guardian) travel to the university in Santiago to work for two days in the detector assembly and data taking. They work in groups of 4 or 5, led by a tutor when assembling the experiment, which takes approximately three to four hours.  We provide all components from Figure~\ref{fig:detectorComponents} as well as gloves, PTFE tape, lubricant, screws, cables and a multimeter per group. An oscilloscope is available in the lab as well, so girls can evidence a muon signal right after they start taking data with their detectors.

\begin{figure}
\includegraphics[width=0.45\textwidth]{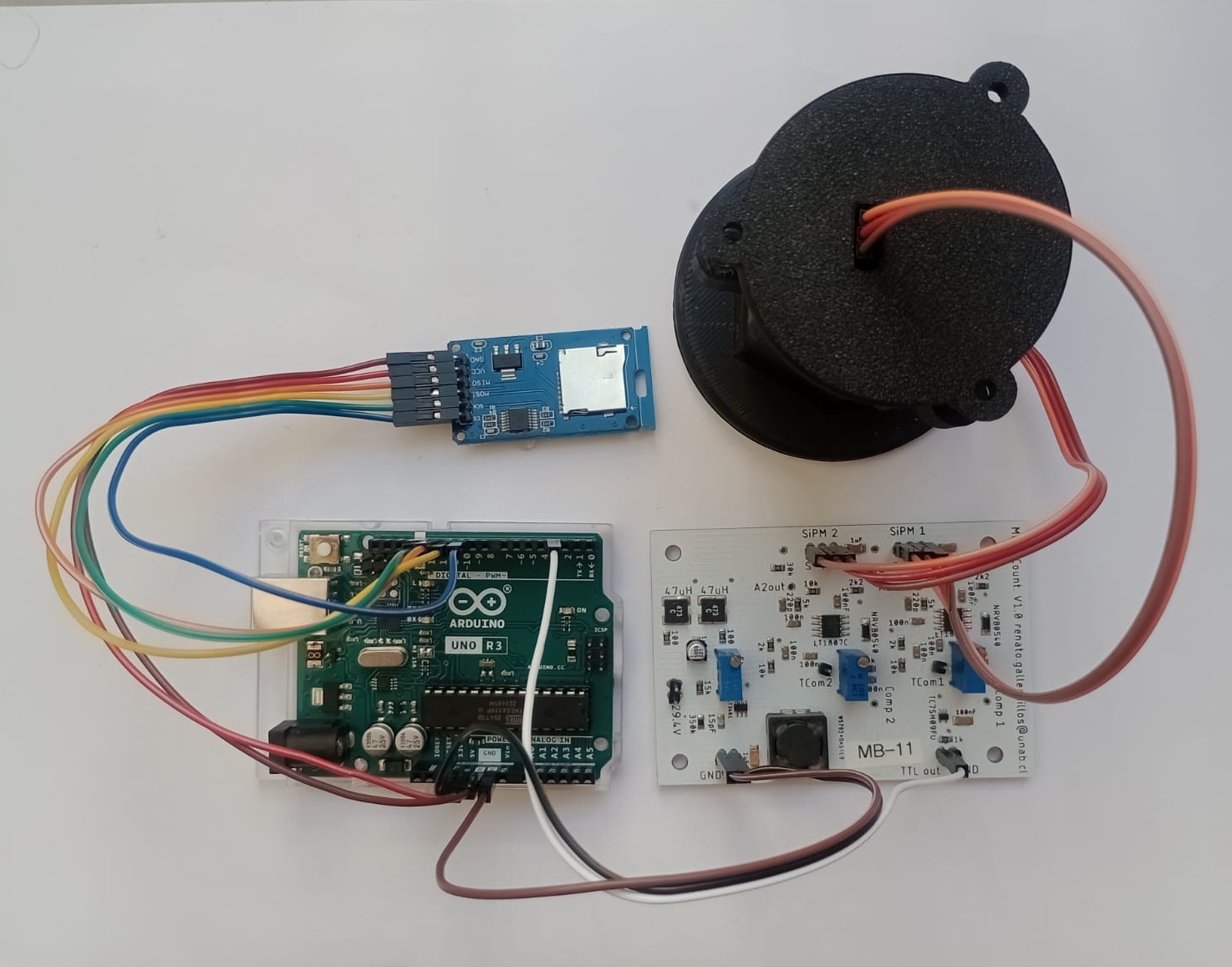}
\caption{Full assembly of our muon detector. The detector components and electrical connections are shown in Figure~\ref{fig:detectorComponents} and \ref{fig:Electronics}, respectively. We thank Ignacio García for taking this photograph.}
\label{fig:fullDetector}
\end{figure}

The detector assembly starts by placing the plastic scintillators and the SiPM inside the detector's hexagonal casings. The students then electrically connect the SiPMs to the PCB, and the PCB to the Arduino. They then supply power to the circuit by connecting the Arduino to a computer with a USB. The electrical connections are shown in Figure~\ref{fig:Electronics}. The full assembly of the experiment is shown in Figure~\ref{fig:fullDetector}. A ``step-by-step" tutorial (in Spanish) made by our tutors, as well as a version in English, can be found in~\cite{AtomicasGithub}. After assembly, the girls install the Arduino IDE Software in their computers, following the instructions in Ref~\cite{Arduino}. The program registers an index for each muon together with the time (in seconds) at which each muon was detected, which the girls can see in real time. A {\texttt{.txt}} file will be saved to the SD card with this data. We detail a specific format for them to name their files specifying their data taking conditions to avoid confusions. When monitoring with a computer that the detectors are working properly, a count of around one muon per minute (at sea level) is expected. We note that the detectors could also be used outside the lab without a computer, by plugging them directly into the mains of a different place using a transformer.

\vspace{0.5cm}

{\bf{Programming:}} Four online sessions of about one hour each are dedicated to explain how they can analyze the data they took with their muon detector. These are distributed in two two-hour modules during the first two days of the second week of the workshop. After each modulus, they have hands-on programming sessions in the afternoon (in breakout rooms over Zoom), rounding up our two days devoted to programming. 

The lessons start with the importance of using programming tools as an agile, clear and streamlined way to communicate with computers and to perform dedicated tasks as analyzing data. We emphasize that learning to program today is as essential as learning a language like English, since it broadens access to knowledge and allows students to draw their own conclusions about natural phenomena, such as counting muons. We provide everyday examples they use as social media, that make use of algorithms written in the language of computers. We also clearly state that the main objective is to plot their different data sets taken under different conditions (e.g. muon fluxes when placing their detectors at different heights or other different conditions they are curious about).

We teach them how to define objects, functions and read files in {\texttt{python}}~\cite{python}. We use Google Colab~\cite{collab} to program and execute a notebook (with a \texttt{.pynb} extension) directly in the browser. Colab and the \texttt{python} notebook provides a shared environment that allows all students to access and run the same application, without worrying about installation issues. We choose {\texttt{python} due to its simplicity and readability, which make it particularly friendly for beginners. } Example {\texttt{python}} script files as well as a Google Colab ``quick-start" tutorial (in Spanish) can be found in~\cite{AtomicasGithub}. We construct these scripts step by step along the programming lessons. 

\vspace{0.5cm}

{\bf{Scientific Methodology: }} Two sessions lasting about one hour (the first in person and the second one online) are devoted to explaining how science is done. The first lecture is given on the second in-person day of the workshop, after the girls assembled and tested their muon detectors (so they can start reflecting early on what to do with their data and under which conditions they will take it). We start the first session with a real story on an anomalous measurement in the time of flight of neutrinos being produced at CERN and then detected by the OPERA experiment in Gran Sasso\footnote{See Ref.~\cite{Opera} for consecutive CERN press releases on the story. See Ref.~\cite{OPERA:2011ijq} for the OPERA measurement (the first preprint on arXiv evidences the anomaly). Also, this particular story on how science works (and others), can also be found in Tim Lewens book in Ref.~\cite{lewens2015meaning}.}. OPERA detected in 2011 an early arrival time of neutrinos when compared to the speed of light in vacuum~\cite{OPERA:2011ijq}. One interpretation of this measurement became popular at that time, that neutrinos could travel faster than the speed of light, breaking the principles of Einstein's special relativity if it were to be true\footnote{We explained earlier in the particle physics sessions what a neutrino is and the importance of special relativity in the measurement of muons.}. Throughout this story, we emphasize three key messages: i) the importance of peer review and validation by the community ii) that the ability to confirm or falsify a result (or even a statement) is necessary if we want to know the truth about something, and that this process requires multiple tests and hard work (sometimes of hundreds of people), as, paraphrasing Carl Sagan, ``extraordinary claims require extraordinary evidence" and that ``only through inquire we can discover truth"~\cite{Sagan1979} iii) the importance of predictive power in a scientific theory, and the need of agreement of these predictions with experiment, considering all experiments to date, no ``matter how beautiful your theory is or how smart you are", quoting Richard Feynman~\cite{Feynman1964}, without invalidating previous knowledge and especially if you don't have enough evidence. In this later message we stress on the value of making predictions or formulating hypothesis, and that this is a brave endeavor. We emphasize that, in the words of Tim Lewens, ``what makes something a genuine piece of science is its potential vulnerability to refutation"~\cite{lewens2015meaning}. We exemplify predictions of both the Standard Model of Particle Physics and General Relativity (e.g. particle masses, how frequently particles are produced by the LHC, the bending of light and gravitational waves), two scientific theories that have passed the tests of time.

Along our second session, we explain how in science we document the process of doing it. We show them examples of real scientific papers, and encourage them to ask themselves a question related to the muons they measured. One of the most common and straightforward question they can ask is : {\it{``Does the muon counts change if I place the detector at different heights?" }}. Another common question is {{\it{``Does the muon counts change if I put some type of material (it could be a rock, or a magnet) on top of the detector?"}}}. The girls then report their findings to such questions individually. They make a scientific report in the following two afternoons after the programming sessions (we also devote the second-to-last day of the workshop for them to meet online with their groups to answer questions on their reports). They can do their reports in any format, including a presentation, an infographic, or a more standard written report, as long as they write it in a conservative and objective language and include the expected contents: i) a clear question or hypothesis they put to the test ii) a brief introduction that provides context and motivates their question iii) an analysis section where they demonstrate what they did and how, including plots or tables with the data they took iv) a summary section, where they must link their findings to their concrete scientific question or hypothesis. They must comment on weather their curiosity can - or can not - be answered with their data. Or to which extent, particularly, if their experiment failed. We encourage them to think on what other possible measurements they could do in order to investigate further to provide answers to their questions, stressing that an important aspect of scientific work is to embrace doubts, and so they can evidence that their work indeed opens more questions.

\vspace{0.5cm}

{\bf{Closure:}} The last day of the workshop starts with dedicated feedback, where each tutor explains what each of the girls in their group did well according to everything we taught them, as well as where there is room for improvement. The main objective of this session is that the girls receive comments of their scientific reports, in a constructive way and in a safe space, as this is done for each group where only their tutors and teammates are present. 
After this, and as a last session of the workshop, we invite a Chilean scientist or an ``Atomic Woman" to give the girls a talk on their work and scientific career. The goal is that the girls can listen to the motivations and unique journey of a female scientist from a different field. Our invited speakers from the 2024 and 2025 workshops gave talks on Data Analysis and Machine Learning, and on Quantum Optics. The workshop ends with an informal ``Q$\&$A" and farewell session. 

\section{Results}
\label{sec:results}

We present combined results using data taken by the girls during the 2024 and 2025 versions of our workshop, with a total of 50 participants. We show in section~\ref{sec:resultsMuons} results on muon flux and proper lifetime, which are two muon properties which can be extracted with the muon detector. In section~\ref{sec:resultsWorkshop}, results from a questionnaire that 40 out of the 50 girls responded after finishing the workshop are also presented, related to the girl's learning perceptions right after completion.

\subsection{Results with the constructed muon detectors}
\label{sec:resultsMuons}

\begin{figure}
\includegraphics[width=0.45\textwidth]{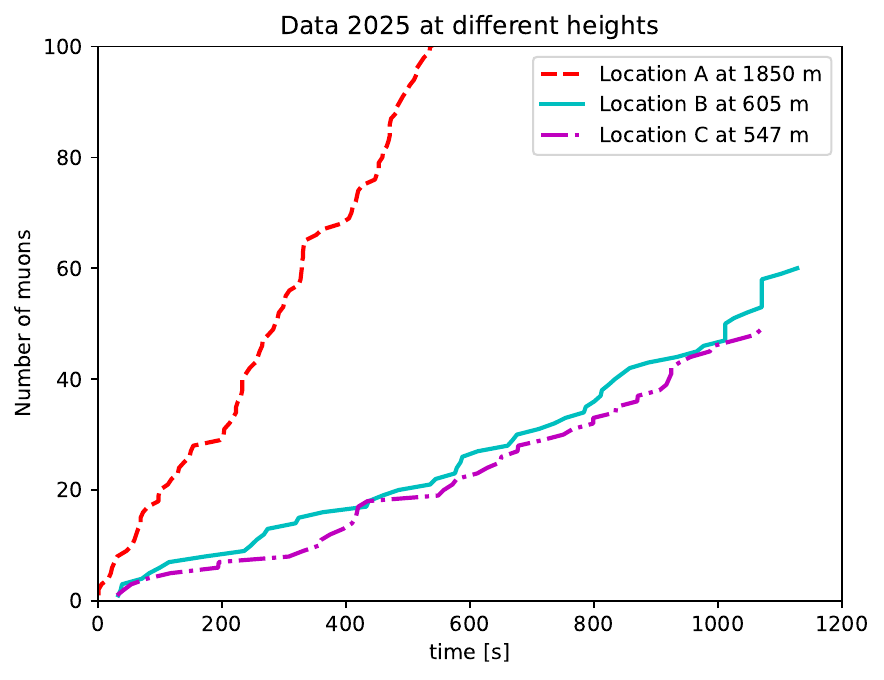}
\caption{Number of muons versus time at different heights. Location A: Yerba Loca park in Farellones near Santiago, at an altitude of 1850 m. Location B: tallest building at San Joaquín campus (605 m). Location C: physics laboratory at San Joaquín campus (547 m).}
\label{fig:heights}
\end{figure}

Figure~\ref{fig:heights} presents the muon data recorded with the muon detectors, showing the number of detected muons as a function of time. The dataset from Location A was collected in the Yerba Loca park in Farellones near Santiago, at an altitude of 1850 m. The remaining measurements were taken at the San Joaquín campus: Location B at the tallest building (605 m) and Location C in our laboratory (547 m). The dataset from Location A is provided for comparison, while the data from Locations B and C correspond to measurements the girls carried out at our university. This plot illustrates a representative result that most of the girls were able to produce using their own measurements during the workshop. By analyzing it they can answer their proposed scientific question:{\it{ ``Does the number of detected muons depend on the altitude at which the data is collected?"}} The answer is yes: with our muon detector this effect is clearly observed. At higher altitudes, the detector registers more muons. However, when comparing only Locations B and C, the differences in altitude are relatively small, so the effect is less pronounced when compared with Location A.

\begin{figure}
\includegraphics[width=0.45\textwidth]{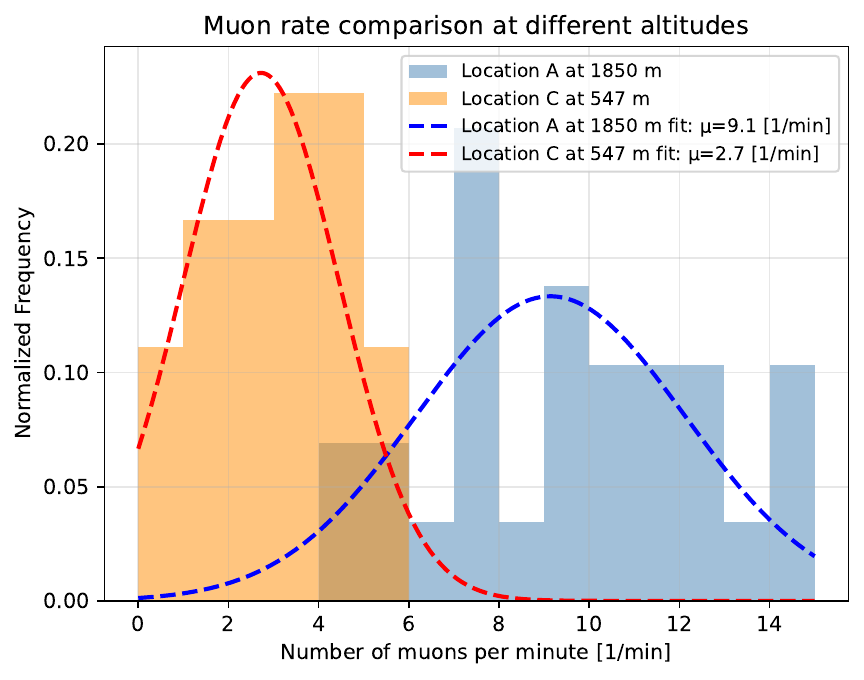}
\caption{Gaussian fits to the data collected at $1850$ m and $547$ m (Locations A and C), where the fitted mean $\mu$ corresponds to the average number of muons per minute detected at that altitude.}
\label{fig:mean_muon}
\end{figure}

In addition, the muon lifetime can, in principle, be extracted from the data collected by the students. However, this has proven to be a very challenging task for them, since explaining the phenomenon requires a precise understanding of concepts from special relativity (which are very unfamiliar to most students). This difficulty is reflected in the fact that none of the participants were able to perform this estimate. Nevertheless, we also include this example to demonstrate the experiment's full potential. Figure~\ref{fig:mean_muon} shows the Gaussian fits to data taken at Location A and Location C. The mean of the distributions corresponds to the average number of muons per minute observed at that altitude. At Location A, we detected a mean of $9.1$ muons per minute. This same exercise can be done at several altitudes. Choosing a second measurement at location C we detected a mean of $2.7$ muons per minute. With these two measurements, we can calculate the muon lifetime.

The number of muons at time $t$ is given by:

\begin{equation}
	N(t) = N_0\exp(-t/\gamma\tau)
\end{equation}

where $N_0$ is the initial amount of muons at a certain height, $\gamma$ is the Lorentz relativistic factor and $\tau$ is the muon lifetime. Therefore, the muon lifetime is given by:

\begin{equation}
	\tau=-t/\ln(N(t)/N_0)\gamma 
\end{equation}

In our example, we assume that the muons travel with a constant velocity of $v_{\mu} = 0.9c$, where $c$ denotes the speed of light, and that the time is given by $t = (h_A - h_C)/v_{\mu}$, with $h_A$ representing the altitude at Location A and $h_C$ the altitude at Location C. For our example, we used $N_0 = 9.1$ and $N(t) = 2.7$, corresponding to the mean number of muons per minute observed at altitudes $h_A = 1850$ m and $h_C = 547$ m, respectively. From these values, we obtain a muon proper lifetime of: 

\begin{equation}
	\tau= 1.7~\text{ns}.
\end{equation}

This is a simple measurement that assumes a constant muon velocity and without taking any uncertainties into account, still it is fairly close to the real value of $2.2~\text{ns}$~\cite{PDGMuonLifetime}. 

Data taking can be further improved. In the 2024 and 2025 editions of the workshop, several measurements were performed by our students over short time intervals of about 20 minutes, in contrast to the data collected at Location A, where the counter operated for approximately one hour. We found that longer acquisition times lead to more stable statistics and more evident Gaussian distributions in the histogram of detected muons per minute, as can bee seen in Figure~\ref{fig:mean_muon}. Therefore, a key lesson learned from this experience is the importance of maximizing the data-taking period of the muon detector whenever possible.

\subsection{Workshop evaluation}
\label{sec:resultsWorkshop}

\begin{figure}[h]

         \centering
         \includegraphics[width=0.49\textwidth]{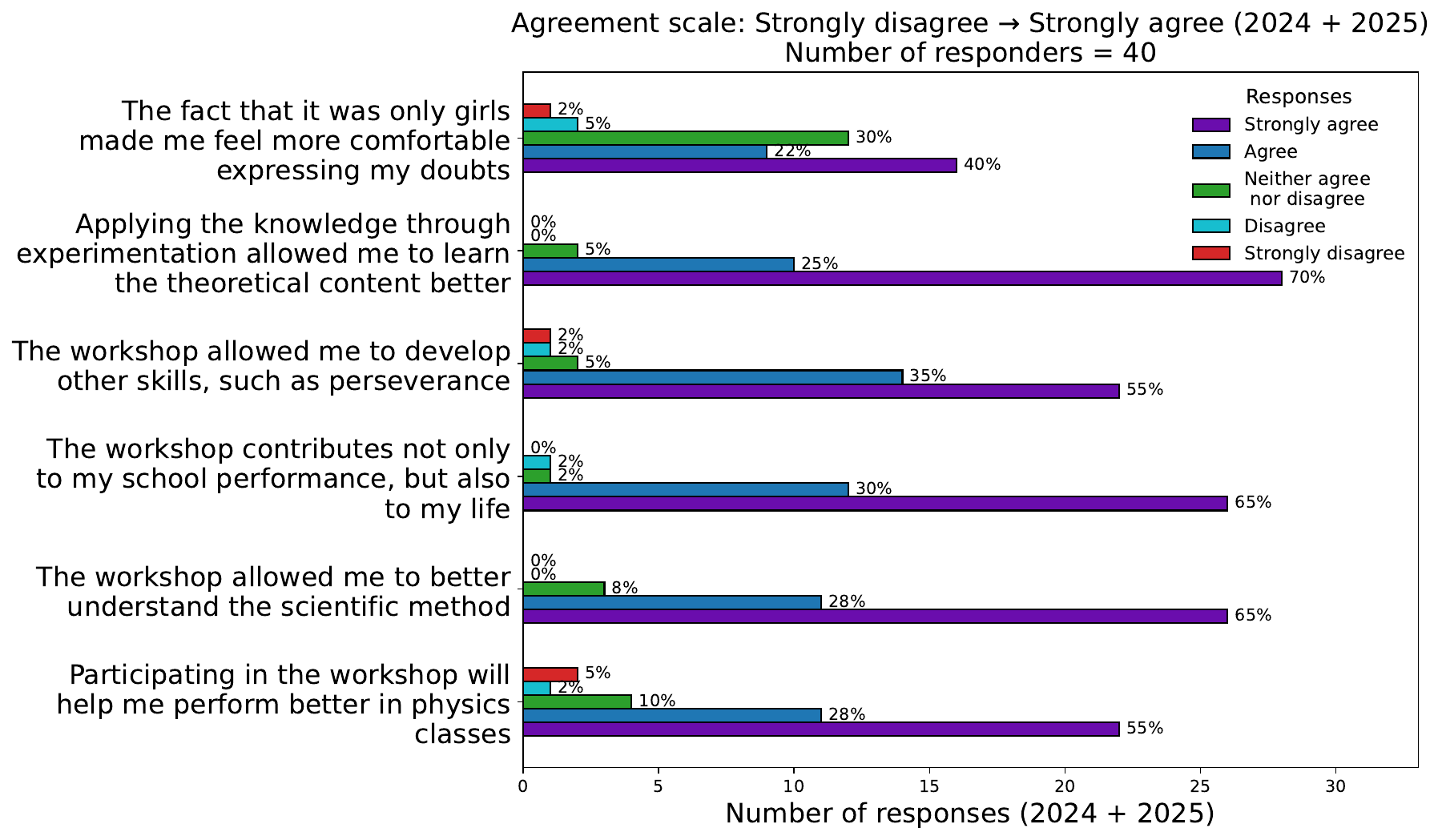}
         \caption{Level of agreement to six statements aligned with our first workshop objective: fostering critical thinking and transferable scientific skills.}
         \label{fig:survey_1}
     \end{figure}

\begin{figure}[h]
         \centering
         \includegraphics[width=0.5\textwidth]{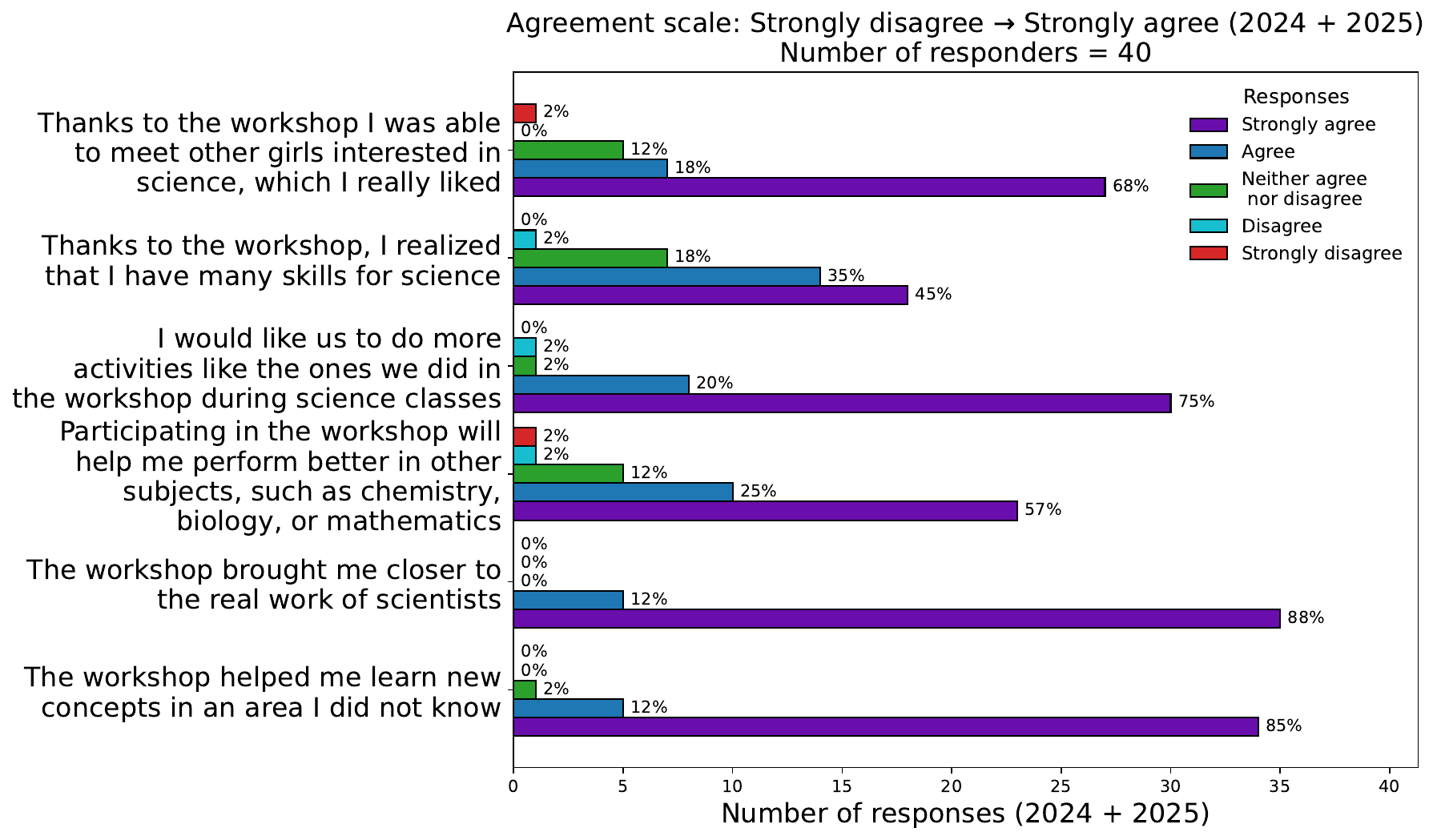}
         \caption{Level of agreement to six statements aligned with our second workshop objective: aid students in their decision-making process regarding a potential STEM career by providing a concrete understanding of what is science and how it is done.}
         \label{fig:survey_2}
     \end{figure}

After each version of the workshop the girls responded to a survey, intended for us to understand better their learning experience and to gather feedback on the workshop methodology and its overall dynamics. This also helps us understand if we are meeting the workshop main objectives (see section~\ref{sec:Intro}). The following results relate to the girls learning perception and workshop appreciation. We present results from the 2024 and 2025 editions of the workshop, which involved a total of 50 participants, of whom 40 completed the survey. Figures~\ref{fig:survey_1} and \ref{fig:survey_2} show the level of agreement of the participants to 12 statements. Statements 1–6 align with our first objective of fostering critical thinking and transferable skills. They address the participants’ perceptions of the abilities they believe they acquired through the workshop, as well as how the workshop contributed to their understanding of science and the learning of new concepts. Statements 7–12 are aligned with our second objective of supporting participants in their decision-making process by providing a concrete understanding of what science is and how it is practiced. These items assess their level of agreement regarding how the workshop influenced their perception of science and the work of scientists, their confidence in current or future performance in STEM subjects, and their views on the workshop being exclusively for girls.

Figure~\ref{fig:survey_1} shows that most participants agreed or strongly agreed that the workshop enhanced their understanding of the scientific method (93\%), helped them learn theoretical content through experimentation (95\%), and contributed not only to their school performance but also to their personal development (95\%). Skills such as perseverance were also highlighted (90\%), and more than half of the girls felt more comfortable expressing doubts in the girls-only setting (62\%).

From figure~\ref{fig:survey_2} we can see that the level of agreement (``agree" or ``strongly agree") to the statements was even stronger: 97\% reported that the workshop helped them learn new concepts, 100\% that it brought them closer to the real work of scientists, and 82\% that it will help them perform better in other science subjects. Moreover, 86\% valued meeting other girls interested in science, 80\% recognized having developing science-related skills, and 95\% expressed the wish to see similar activities integrated into regular science classes. These results suggest that the workshop not only fostered transferable skills but also shaped participants perceptions of science, scientists, and their own potential in STEM.

\begin{figure}[h]
\includegraphics[width=0.45\textwidth]{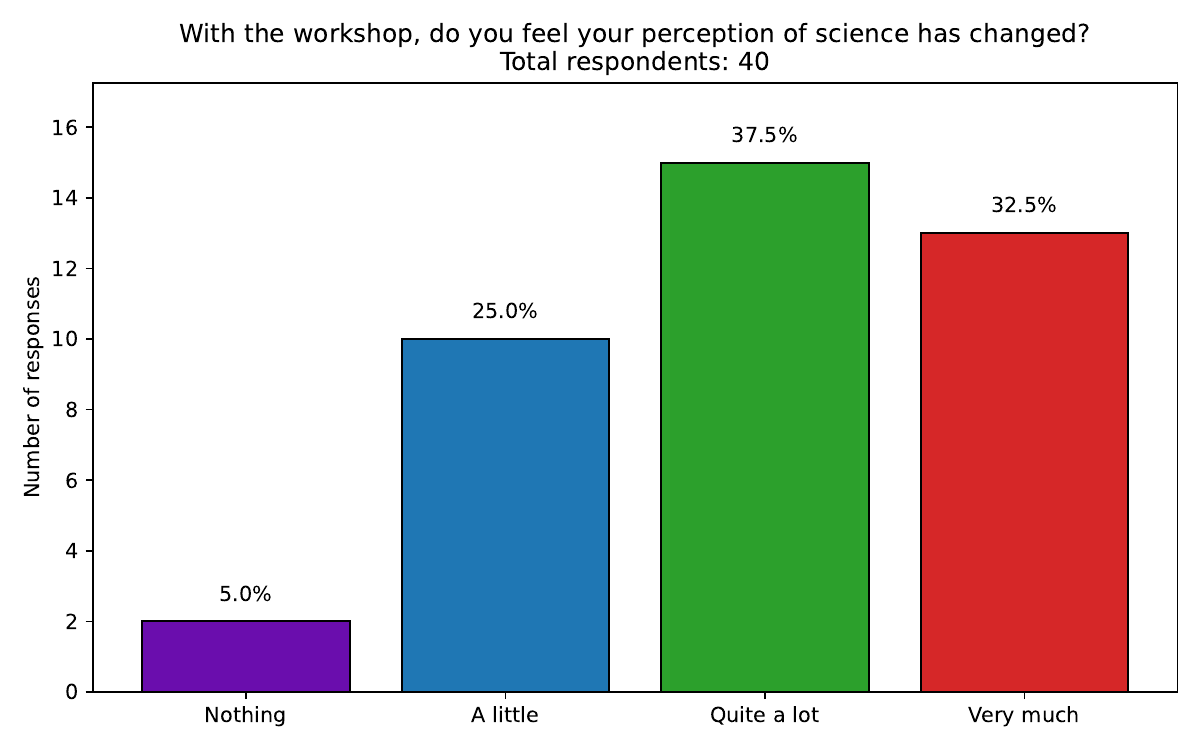}
\caption{Perception level on science in a scale of 4 after finishing the workshop.}
\label{fig:MI2}
\end{figure}

\begin{figure}[h]
\includegraphics[width=0.45\textwidth]{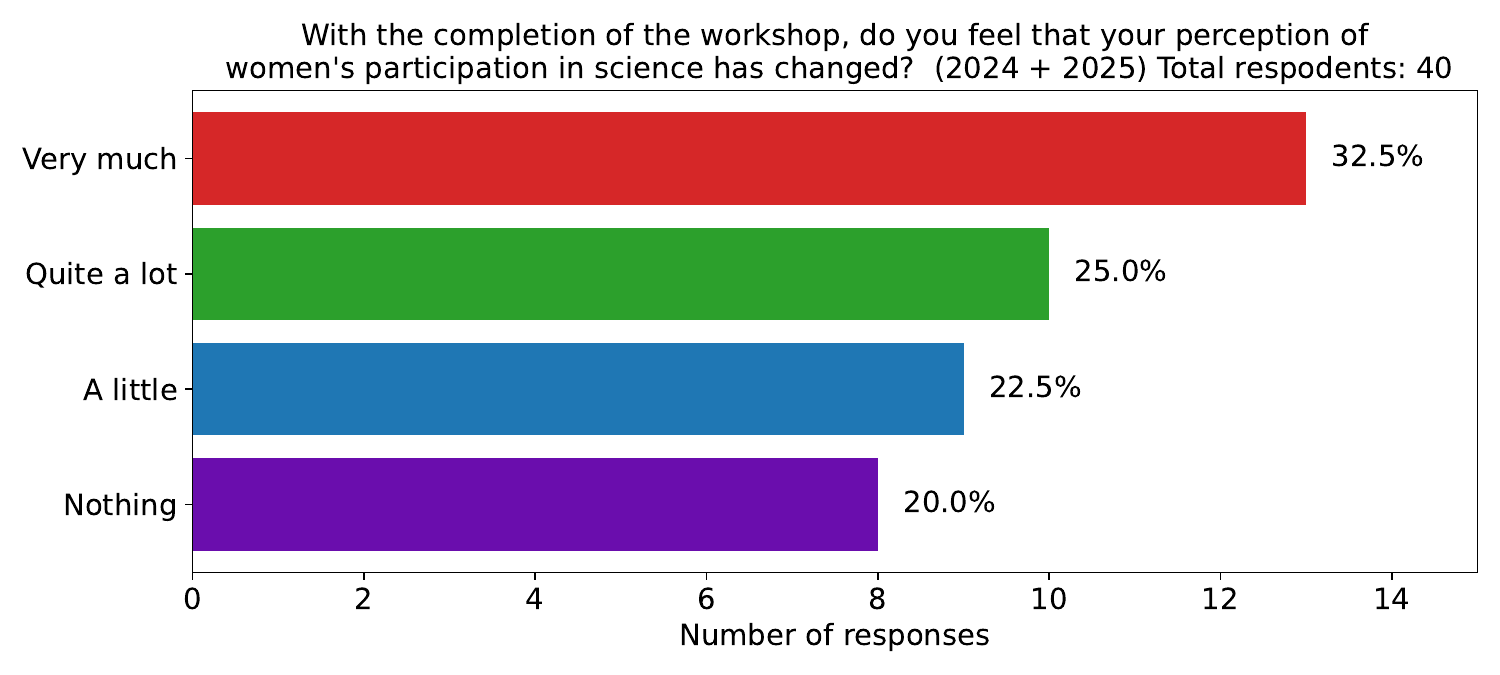}

\caption{Perception level on women's participation in science in a scale of 4 after finishing the workshop.}
\label{fig:MI3}
\end{figure}

In addition to evaluating specific skills, the survey also asked participants to reflect on how the workshop influenced their views on science and on women’s participation in science. As shown in figure~\ref{fig:MI2}, a 70\% of the girls reported that their perception of science changed ``quite a lot" or ``very much", while only 5\% indicated no change at all. In figure~\ref{fig:MI3} we see that a 57.5\% of respondents indicated that their perceptions concerning the participation of women in science changed ``quite a lot" or ``very much", while 20\% responded ``nothing" . Even though we do not present results at this time on what are precisely their previous perceptions prior to the workshop (another work is in progress in~\cite{EstudioImpacto}), these responses suggests that most of the participants are, at the very least, {\it{questioning}} their perceptions after participating in the workshop.

\section{Summary and outlook}
\label{sec:summary}

We have created an initiative that combines the teaching of particle physics, programming, electronics and scientific methodology through the building of a dedicated particle physics experiment: a muon detector that counts muons. Our ``Niñas Atómicas" workshop has run for two weeks every year since 2022, where high-school girls aged 15 to 16 years old from all over Chile can experience how  science works. Our main objectives are to foster transferable scientific skills as well as to support students in their decision-making process when considering a scientific career. 

The usage of ``hands-on" physics experiments provides an engaging way to introduce students to complex concepts. Our workshop revolves around building a muon detector, which is a portable particle detector that the girls use to explore fundamental physics phenomena. Several experiments using muon detectors, intended for educational use at various levels, are also detailed in~\cite{Coan_2006,Singh2015MuonDetection, Barazandeh2016MuonTeaching,Axani_2017,Bosnar:2018ngm,Axani:2018qzs}. 
Simple experimental setups targeted at school-level students for cosmic muon lifetime measurements, relying also on scintillators, are described in~\cite{Singh2015MuonDetection, Bosnar:2018ngm}. 
Compact and low cost muons detectors for undergraduate-level students can also be found in the literature. The design in~\cite{Coan_2006} allows for in-field lifetime and time dilation measurements. With the experiment in Ref.~\cite{Barazandeh2016MuonTeaching}, muon lifetime as well as the energy spectra of muons can also be collected. A more detailed experimental setup is presented in~\cite{Axani_2017}, with which muon rate measurements as a function of the polar angle can be done with two of these detectors. A more complex pocket-sized muon detector design aimed for outreach activities, with a built-in screen that can display the muon count, is also described in~\cite{Axani:2018qzs}. All these muon detector setups could be used in initiatives of this kind. 

Other initiatives that include the use of interactive physics apparatus at schools suggest an increased interest in the pursuit of a STEM career~\cite{Lubrica2017}. Teaching particle physics at the high-school level is also a valuable experience for diversifying students backgrounds in physics before entering the university~\cite{RogerBarlow_1992}, and have had a long-term impact worldwide, for instance, with the development of the Particle Physics Masterclasses~\cite{Masterclasses,Bilow:2022rs,2024arXiv240117103T}. Moreover, initiatives of this kind can aid bring young people awareness on High Energy Physics (HEP) in Latinamerica~\cite{Assamagan:2023eac}.

Some key challenges we have faced are worth noting. The amount of logistics and funding needed to develop and materialize an initiative of this kind required strong and constant support from funding agencies and universities. We performed the workshop in hybrid mode twice, and girls from many regions of Chile participated. This demanded funding for both the girls and their legal guardians to stay in Santiago for a few days. In addition, the estimated cost of each muon detector is close to 300 USD.
These take important budget considerations when thinking of performing and/or scaling this initiative.

Other aspect to consider is that the girls must have access to a computer with an internet connection. We do require this at the time of application, so their effective participation relies on them to be able to access a computer (not cellphones or tablets) during the remote sessions at their homes, high schools or local communities. 
An additional challenge for us was a limited amount of reliable resources on particle physics in Spanish. This meant to dedicate particular efforts in making our lessons and material as self-contained as possible. As most of the girls that participated do not speak well English, and most science is done and communicated in English, we aim to contribute to narrow this gap in resources with time. A detector assembly manual and particle physics handbook created by our tutors can be found in~\cite{AtomicasGithub}.

As the participants of the workshop are minors, many legal, ethical and administrative efforts had also to be considered\footnote{Some of these challenges are also described in other physics and programming outreach programs~\cite{Briceno:2024nfi}.}, which were very new and time consuming for us scientists that usually work with data with no ethical concerns, such as the one coming from proton-proton collisions at the Large Hadron Collider at CERN.

None of the above challenges were limiting factors in the creation of this initiative, which we hope remains stable in time. Our overarching goal is to cultivate scientific thinking among young minds in Chile, enabling them to become ambassadors at their local communities as representatives of our ``Niñas Atómicas" program. The final scientific reports on their work at the workshop have become a valuable resource, as the girls themselves can share them and present them at their high schools and local scientific fairs.
We hope to keep encouraging students to ask questions, develop the tools and skills to be able to answer them, and more importantly, to inspire them to remain curious about the universe we live in.

\medskip

\subsection*{Acknowledgments}
We thank Renato Galleguillos and the Center for Theoretical and Experimental Particle Physics
(CTEPP) at Universidad Andrés Bello (UNAB) for the design of the muon detector and the PCB, which is essential for this muon experiment. We thank Florencia Díaz (M.Sc. student in Physics, UC), our electronics teacher and Suyay Huichacura  (M.Sc. student in Medical Physics, UC), Melanie M. Villarreal (Ph.D. (c) in Physics, UC), Devika Mukhi-Nilo (M.Sc. student in Astrophysics, UC) and Mariel Poduje (Ph.D. (c) in Physics, UC), our senior tutors of the ``Niñas Atómicas" program for guiding the students and aiding them in the data taking, as well as for many useful discussions concerning the workshop over the past years. In particular, we thank Mariel Poduje and Devika Mukhi-Nilo for the Spanish version of the manual. We are grateful to Gabriel Torrealba, Florencia Díaz and Devika Mukhi-Nilo for useful comments on this manuscript. We thank Ignacio García for taking updated photographs of the experiment and his review of the detector manual and detailed instructions in English. We also acknowledge our administrative and logistics support team at SAPHIR Millennium Institute.  We thank Roberto Pinto for his constant support of our initiative and the Institute of Physics at the Faculty of Physics of Pontificia Universidad Católica de Chile for providing laboratory space and financial support in the developing of the 2024 and 2025 versions of the workshop. We also acknowledge financial support from ANID FONDECYT grant No. 1240721 and ANID – Millennium Science Initiative Program ICN2019\_044, as well as the Vice-Rectorate of Research at Pontificia Universidad Católica de Chile.

\medskip

\subsection*{Data availability statement}
The data that support the findings of this study regarding the responses of the participants of the workshop are included within the article. Data taken with the muon detectors can be found in~\cite{AtomicasGithub}. 

\subsection*{Ethical statement}

This study was approved by the Vice-Rectorate of Research at Pontificia Universidad Católica de Chile and by the program ``Proyección al Medio Externo" (PME) by ANID Millennium Science Initiative Program. All parents, guardians or next of kin provided written informed consent for the minors to participate in this study and we have consent for the student’s results to be published.

\bibliography{bibitem}
\bibliographystyle{apsrev4-1}

\end{document}